\documentclass[10pt,aps,prl,twocolumn,superscriptaddress]{revtex4}

\usepackage{hyperref}
\usepackage{graphicx}
\usepackage{amsmath,amssymb}
\usepackage{subfigure}
\usepackage{url}
\usepackage{flushend}

\graphicspath{{Images/}}

\begin{document}

\title{
Absolute measurement of small-amplitude vibrations by time-averaged heterodyne holography with a dual local oscillator
}
\author{Nicolas Verrier}
\author{Michael Atlan}


\affiliation{
Institut Langevin. Fondation Pierre-Gilles de Gennes. Centre National de la Recherche Scientifique (CNRS) UMR 7587, Institut National de la Sant\'e et de la Recherche M\'edicale (INSERM) U 979, Universit\'e Pierre et Marie Curie (UPMC), Universit\'e Paris 7. \'Ecole Sup\'erieure de Physique et de Chimie Industrielles - 10 rue Vauquelin. 75005 Paris. France\\
}
\date{\today}
\begin{abstract}
We report a demonstration of the measurement of the ratio between  an optical modulation side band component and the non-shifted light component by time-averaged heterodyne holography in off-axis and frequency-shifting configuration, through coherent frequency-division multiplexing with a dual optical local oscillator. Experimental results of sinusoidal vibration sensing are reported. This technique enables absolute measurements of sub-nanometric out-of-plane vibration amplitudes. 
\end{abstract}
\maketitle

Laser Doppler interferometric methods are commonly used for non-contact measurements of mechanical vibrations. These methods exhibit high reliability and enable wideband, phase-resolved, single point vibration measurements \cite{BarriereRoyer2001}. However, imaging requires time-consuming scanning of the tested sample. Homodyne \cite{Powell1965, Picart2003, Pedrini2006} and heterodyne \cite{Aleksoff1971, UedaMiida1976, JoudLaloe2009} holographic recordings in off-axis configuration enabled reliable full-field measurements of out-of-plane mechanical vibrations. Nevertheless, quantitative measurements of vibration amplitudes much smaller than the optical wavelength with an array detector remains difficult to achieve. A comprehensive study of the signal-to-noise ratio (SNR) was proposed for classical heterodyne holography \cite{UedaMiida1976}. The authors managed to observe vibration amplitudes of a few Angtroms, and linked the smallest detectable amplitude to the SNR, in the absence of spurious effects. Later on, nanometric vibration amplitude measurements were achieved with digital holography, by sequential measures of the first optical side band and the non-shifted light component\cite{PsotaLedl2012}.\\

In this letter, we report an experimental demonstration of heterodyne holography for vibration sensing. The presented idea is to make use of a dual optical local oscillator (LO) illumination to assess low-amplitude modulations. Optical heterodyning is a frequency-conversion process used to shift the radiofrequency (RF) content of an optical radiation field $E$ in a sensor bandwidth by mixing it with a LO field $E_{\rm LO}$ whose optical frequency is detuned by the RF of interest. The RF spectrum of an optical beam undergoing sinusoidal modulation exhibits modulation side bands. By combining two LO, both the fundamental and the first-order harmonic optical side bands can be recorded simultaneously. This approach has its roots in frequency-division multiplexing \cite{Weinstein1971}, a technique by which the total bandwidth available is divided into non-overlapping frequency sub-bands, each of which is used to carry a separate signal. The proposed method is shown to be suitable for the detection of the complex ratio of modulated light side bands.\\

\begin{figure}[t]
\centering
\includegraphics[width = 8 cm]{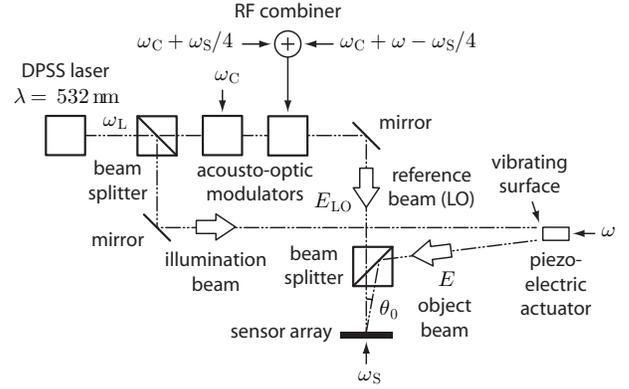}
\caption{Sketch of the acquisition setup.}
\label{fig_setup}
\end{figure}
\begin{figure}[t]
\centering
\includegraphics[width = 8.0 cm]{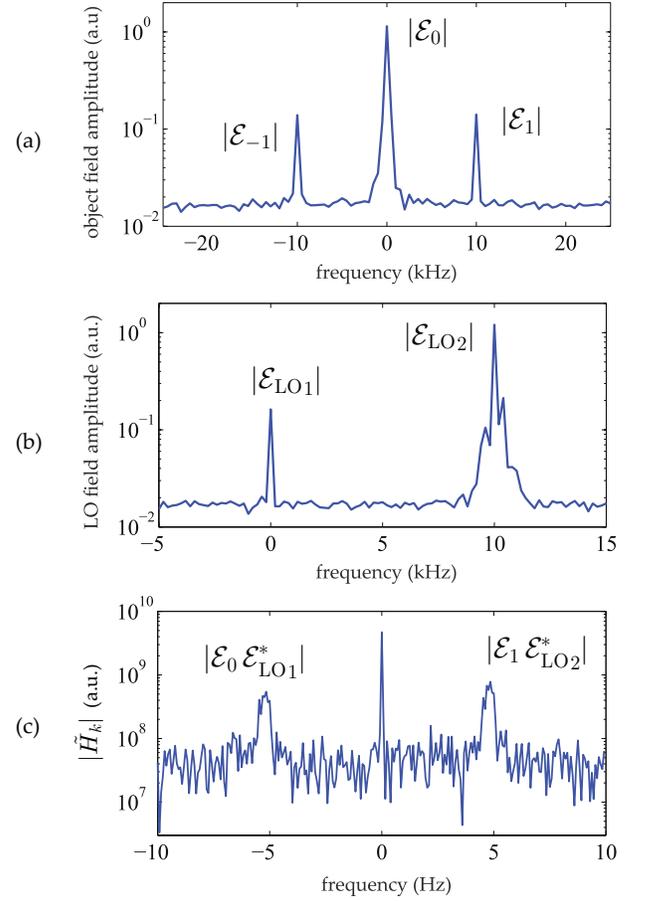}
\caption{(a) Magnitude of the spectrum of the object field $E$ whose phase is modulated at 10 kHz, showing the first lateral bands $\propto |E_{\pm 1}|$ and the non-shifted component $\propto |E_{0}|$ (supply voltage : 1 V) . (b) Magnitude of the spectrum of the dual local oscillator $E_{\rm{LO}} = E_{\rm{LO}_1} + E_{\rm{LO}_2}$. (c) Magnitude of the discrete Fourier transform  $\tilde{H}$, showing the bands of order 0 and +1 shifted within the temporal bandwidth of the camera.}
\label{fig_spectra}
\end{figure}
\begin{figure}[t]
\centering
\includegraphics[width = 8 cm]{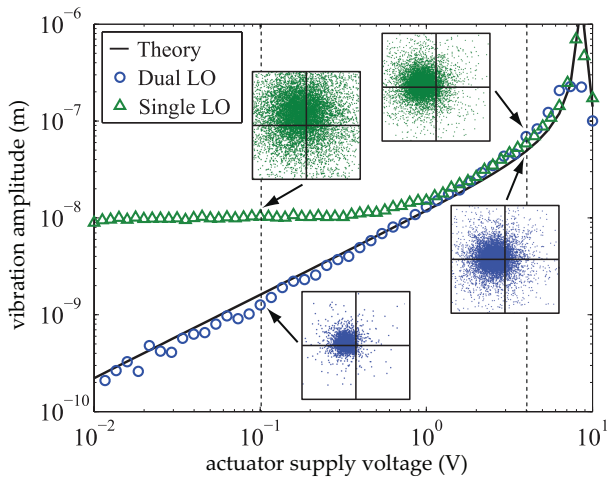}
\caption{Modulation amplitude versus PZT supply voltage : theoretical curve (line), sequential single LO measurement (triangles), dual LO measurement (circles). Insets : complex constellation diagrams of $\tilde{H}\left({\omega_{\rm{S}}/4}\right)/{\tilde{H}\left({-\omega_{\rm{S}}/4}\right)}$ at 0.1 V and 4 V with both approaches; axis ranges at a given supply voltage are the same.}
\label{fig_z_vs_V}
\end{figure}

The acquisition set-up, sketched in fig. \ref{fig_setup} and described in \cite{SamsonVerpillat2011}, consists of an off-axis, frequency-shifting Mach-Zehnder interferometer designed for heterodyne detection of optical modulation side bands. The sensor array used for detection is an EMCCD camera, (Andor IXON 885+, readout rate $\omega_{\rm S} /(2\pi) = 20 \, \rm Hz$). The main optical radiation field is provided by a doubled Nd:YAG laser (Oxxius SLIM 532, power 100 mW, wavelength $\lambda = 532$ nm, optical pulsation $\omega_{\rm L}$). As a result of optical pathlength modulation of the laser beam impinging on the vibrating surface of a piezo-electric actuator (PZT, Thorlabs AE0505D08F), the temporal part of the object field $E$ undergoes a sinusoidal phase modulation of the form $\phi(t) = \phi_{0} \sin(\omega t)$, where $\omega/(2\pi) = 10 \, \rm kHz$ is the excitation frequency. It can be decomposed on a basis of Bessel functions of the first kind $J_n(\phi_0)$, via the Jacobi-Anger identity
\begin{equation}
E = \sum_{n=-\infty}^{\infty} E_n = \sum_{n=-\infty}^{\infty} {\cal E}_n e^{i\left(\omega_{\rm{L}}+n\omega\right)t}
\label{eq_JacobiAngerDecomposition}
\end{equation}
The phase modulation of the object field $E$ at angular frequency $\omega$ results in the apparition of optical side bands of complex amplitude ${\cal E}_n$ at harmonics of $\omega$; for small modulation depths, the magnitude of the side bands of order $\pm 1$ is much greater than the magnitude of the side bands of higher order, as reported in figure \ref{fig_spectra}(a). The quantity
\begin{equation}
{\cal E}_n = {\cal E} J_n \left(\phi_0\right)
\label{eq_Complexside bandsWeights}
\end{equation}
is the complex weight of the optical side band of order $n$, where ${\cal E}$ is the complex amplitude of the optical field, and $\phi_0 = 4\pi z/\lambda$ is the modulation depth of the optical phase. For small vibrations ($z \ll \lambda$), a relative measure of $z$ can be assessed from the first-order side band hologram $z \propto | {\cal E}_1 |$ \cite{SamsonVerpillat2011}. Furthermore, the local amplitude $z$ of the out-of-plane motion at angular frequency $\omega$ is approximately
\begin{equation}
z \approx \frac{\lambda}{2\pi}\frac{J_1\left(\phi_0\right)}{J_{0}\left(\phi_0\right)}.
\label{eq_z}
\end{equation}
Hence, a quantitative measure of $z$ can be achieved by forming the ratio between the the magnitude of the weights of the first-order side band hologram $\propto  {\cal E}_1 $ and the non-shifted light component hologram $\propto  {\cal E}_0 $, each of them being measured sequentially \cite{LedlVaclavik2010}. However, in experimental conditions, a noise floor prevents accurate assessment of $z$ values below 10 nm (Fig. \ref{fig_z_vs_V}) from sequential measures of holograms and spatial averaging of the quantity ${\cal E}_1 / {\cal E}_0$ over the whole image of the piezo-electric actuator (triangles). The sensitivity of the measurement of $z$ can be further enhanced by spatial averaging of the complex-valued ratio ${\cal E}_1 / {\cal E}_0$, if the first-order side band hologram $\propto {\cal E}_1$ and the non-shifted light component hologram $\propto {\cal E}_0$ are acquired simultaneously (Fig. \ref{fig_z_vs_V}, circles). Simultaneous measurement of side bands holograms at both optical modulation bands can be performed by a rudimentary coherent frequency-division multiplexing scheme with a dual LO, which will shift ${\cal E}_0$ and ${\cal E}_1$ in the available temporal bandwidth of the camera, ensuring phase-matching of these quantities.\\

The LO signal consists of the addition of two coherent (phase-locked) RF signals, shifted by a carrier frequency $\omega_{\rm C}/(2\pi) \sim 80\, \rm  MHz$ set around the peak frequency response of acousto-optic modulators used to shift the optical frequency of the laser beam. This summation is done in practice with a power splitter/combiner (figure\ref{fig_setup}), resulting in a LO field of the form $E_{\rm{LO}} = E_{\rm{LO}_1} + E_{\rm{LO}_2}$, with 
\begin{eqnarray}
E_{\rm{LO}_1}& = & {\cal E}_{\rm{LO}_1} \exp{\left[i\left(\omega_{\rm{L}}+\omega_{\rm{S}}/4\right)t\right]}\\
E_{\rm{LO}_2} & = & {\cal E}_{\rm{LO}_2} \exp{\left[i\left(\omega_{\rm{L}} + \omega - \omega_{\rm{S}}/4\right)t\right]}
\end{eqnarray}
where ${\cal E}_{\rm{LO}_1} = \alpha {\cal E}_{\rm{LO}}$ and ${\cal E}_{\rm{LO}_2} = \beta {\cal E}_{\rm{LO}}$ are the complex magnitudes of the LO components. The positive parameters $\alpha$ and $\beta$ (which satisfy the relation $\alpha + \beta = 1$) are the normalized relative weights of each LO component, whose magnitudes are reported in fig. \ref{fig_spectra}(b). In the reported experiment, $\alpha/\beta = 1/10$. In these conditions, the interferogram impinging on the sensor array has the form
\begin{equation}\label{eq_interferogram}
{\cal I} = \left| \textstyle{\sum_{n}} E_{n} + E_{\rm{LO}_1} + E_{\rm{LO}_2} \right|^2
\end{equation}
From which only three contributions are within the sensor temporal bandwidth, between the Nyquist frequencies $\pm \omega_{\rm S} / 2$. The part of the off-axis hologram $H$ modulated at frequencies within the sensor bandwidth is
\begin{equation}\label{eq_interferogram_in_CCD_BW}
H(t) = {\cal E}_{0}{\cal E}^{*}_{\rm{LO}_1} {e}^{-i \omega_{\rm{S}}t / 4}
+ {\cal E}_{1}{\cal E}^{*}_{\rm{LO}_2} {e}^{i \omega_{\rm{S}}t / 4}
\end{equation}
from which a remaining static contribution, which yields the DC peak in figure \ref{fig_spectra}(c), is neglected. The two terms in the right member of eq. \ref{eq_interferogram_in_CCD_BW} yield the peaks proportional to $|{\cal E}_{0}{\cal E}^{*}_{\rm{LO}_1}|$ and $|{\cal E}_{1}{\cal E}^{*}_{\rm{LO}_2}|$ in figure \ref{fig_spectra}(c). We sought to measure the modulation amplitude $z$ of the actuator oscillating at $\omega / (2 \pi) = 10 \, \rm kHz$, at low supply voltages ranging from $10^{-2}$ V to 10 V. For each voltage, $N = 256$ interferograms ${\cal I}_p, \, \{p=1, \ldots N \}$ were acquired. Each recorded interferogram ${\cal I}_p$ was turned into a complex-valued hologram $I_p$ of the PZT surface by a numerical Fresnel transform \cite{VerrierAtlan2011}. The off-axis region of  $I_p$, noted $H_p = H(2 \pi p / \omega_{\rm S})$, was Fourier-transformed temporally; the $k$-th point of the discrete transform is
\begin{equation}
\tilde{H}\left(\omega_k\right) = \textstyle{\sum_{p=1}^N} H_p \exp{\left(- 2 i p k \pi / N \right)}
\label{eq_demod}
\end{equation}
where the $\omega_k$ are linearly spaced between the Nyquist frequencies $-\omega_{\rm S}/2$ and $\omega_{\rm S}/2$. The spectrum $|\tilde{H}\left(\omega_k\right)|$, reported in figure \ref{fig_spectra}(c), exhibits two peaks at $\omega_k = \pm \omega_{\rm{S}}/4$. As expected from the relation \ref{eq_interferogram_in_CCD_BW}, the magnitude of the peak at $-\omega_{\rm{S}}/4$ is proportional to the magnitude of $\alpha {\cal E}^*_{\rm LO}{\cal E}_{0}$, and the magnitude of the peak at $\omega_{\rm{S}}/4$ is proportional to $\beta {\cal E}^*_{\rm LO}{\cal E}_{1}$. Vibration amplitudes $z$ were then calculated from the relation
\begin{equation}\label{eq_z_measure}
	z \approx \frac{\lambda}{2\pi} \frac{\alpha}{\beta} \left|\left\langle {\tilde{H}\left({\omega_{\rm{S}}/4}\right)}/{\tilde{H}\left({-\omega_{\rm{S}}/4}\right)} \right\rangle \right|
\end{equation}
where $\langle \, \rangle$ corresponds to spatial averaging over $100 \times 100$ pixels. For each voltage, sequential single LO measurements from 256 raw interferograms per band  and simultaneous, dual LO measurements from 256 raw interferograms are used to calculate $z$ values from the relation \ref{eq_z_measure}, which are reported in Fig. \ref{fig_z_vs_V}. The theoretical evolution of $z$, according to eq. \ref{eq_z}, for a modulation depth that scales up linearly with the voltage, was added as a guide to the eye. The dual LO approach enables assessement of sub-nanometer vibration amplitudes, in agreement with to the theoretical evolution of $J_1/J_0$ (line), at least one order of magnitude below sequential single LO measurements, for which the noise floor lies at $z \sim 10 \, \rm nm$, in the same experimental conditions. The benefits of the combination of two LO signals arise for small-amplitude vibrations. The phase and the amplitude of the complex-valued quantities ${\cal E}_1/{\cal E}_0$, at low signal levels are measured with much better sensitivity in dual LO regime than with the single LO approach. In particular, the phase relation of $\tilde{H}\left({\omega_{\rm{S}}/4}\right)/{\tilde{H}\left({-\omega_{\rm{S}}/4}\right)}$ between two pixels should only depend on the relative local phase retardation of the sinusoidal motion between those pixels in an ideal measurement. As it can be seen in the constellation diagrams in the complex plane (fig. \ref{fig_z_vs_V}, insets), the complex values of $\tilde{H}\left({\omega_{\rm{S}}/4}\right)/{\tilde{H}\left({-\omega_{\rm{S}}/4}\right)}$ are much less dispersed in dual LO regime than for the sequential, single LO measure. Since all the points of the actuator oscillate in phase, it enabled coherent averaging over the extent of the actuator's image (eq. \ref{eq_z_measure}), which improved the SNR at low modulation depths.\\

In conclusion, we have proposed a robust coherent frequency-division multiplexing method to perform absolute measurements of small-amplitude sinusoidal optical phase modulation by time-averaged heterodyne holography in off-axis and frequency-shifting conditions. The scheme was validated by a quantitative measurement of sub-nanometric vibration amplitudes of a piezo-electric actuator. A dual optical local oscillator was introduced to shift and record two optical modulation bands simultaneously in the temporal bandwidth of the detector array. This approach enabled a measurement of the ratio of the complex weights of the optical side bands with increased sensitivity with respect to a sequential measurement of the two bands, performed in the same experimental conditions.\\


\bibliographystyle{ol}


\begin{thebibliography}{10}
\newcommand{\enquote}[1]{``#1''}

\bibitem{BarriereRoyer2001}
C.~Barriere and D.~Royer, Applied Physics Letters \textbf{79}, 878 (2001).

\bibitem{Powell1965}
R.~L. Powell and K.~A. Stetson, J. Opt. Soc. Am. \textbf{55}, 1593 (1965).

\bibitem{Picart2003}
P.~Picart, E.~Moisson, and D.~Mounier, Applied Optics \textbf{42}, 1947 (2003).

\bibitem{Pedrini2006}
G.~Pedrini, W.~Osten, and M.~E. Gusev, Appl. Opt. \textbf{45}, 3456 (2006).

\bibitem{Aleksoff1971}
C.~C. Aleksoff, Applied Optics \textbf{10}, 1329 (1971).

\bibitem{UedaMiida1976}
M.~Ueda, S.~Miida, and T.~Sato, Applied Optics \textbf{15}, 2690 (1976).

\bibitem{JoudLaloe2009}
F.~Joud, F.~Lalo\"{e}, M.~Atlan, J.~Hare, and M.~Gross, Opt. Express
  \textbf{17}, 2774 (2009).

\bibitem{PsotaLedl2012}
P.~Psota, V.~Ledl, R.~Dolecek, J.~Erhart, and V.~Kopecky, Ultrasonics,
  Ferroelectrics and Frequency Control, IEEE Transactions on \textbf{59}, 1962
  (2012).

\bibitem{Weinstein1971}
S.~Weinstein and P.~Ebert, Communication Technology, IEEE Transactions on
  \textbf{19}, 628 (1971).

\bibitem{SamsonVerpillat2011}
B.~Samson, F.~Verpillat, M.~Gross, and M.~Atlan, Opt. Lett. \textbf{36}, 1449
  (2011).

\bibitem{LedlVaclavik2010}
V.~Ledl, J.~Vaclavik, R.~Dolecek, and V.~Kopecky, in \emph{9th International
  conference on vibration measurements by laser and non contact techniques
  (AIVELA)} (2010).

\bibitem{VerrierAtlan2011}
N.~Verrier and M.~Atlan, Appl. Opt. \textbf{50}, H136 (2011).

\end{thebibliography}

\end{document}